\documentclass[10pt, conference]{IEEEtran}
\IEEEoverridecommandlockouts
\usepackage{cite}
\usepackage{amsmath,amssymb,amsfonts}
\usepackage{algorithmic}
\usepackage{graphicx}
\usepackage{textcomp}
\usepackage{xcolor}
\usepackage{comment}
\usepackage{booktabs}
\usepackage{multirow}
\usepackage{soul,color}
\usepackage{subcaption}
\usepackage{url}
\usepackage{balance}
\def\BibTeX{{\rm B\kern-.05em{\sc i\kern-.025em b}\kern-.08em
    T\kern-.1667em\lower.7ex\hbox{E}\kern-.125emX}}
\begin{document}

\title{Lightweight Replay Attack Detection via Interpretable Physical Descriptors}

\author{\IEEEauthorblockN{Saba Nasiri$^{a, b}$, Alireza Amirshahi$^{c}$, Milos Cernak$^{b}$}
\IEEEauthorblockA{\textit{$^{a}$EPFL, $^{b}$Logitech Audio, $^{c}$Logitech Webcam} \\
Lausanne, Switzerland}}



\maketitle

\begin{abstract}
This paper proposes a lightweight, interpretable audio anti-spoofing method that distinguishes between bona fide human speech and replay attacks by analyzing physical acoustic signatures rather than learned spectro-temporal patterns. Unlike ``black-box'' Deep Neural Networks (DNNs), our approach exploits fundamental physical invariants, specifically source kinematics, binaural phase dynamics, and aerodynamic consistency, captured via multi-channel microphone arrays. By extracting these explicit physical descriptors and classifying them with an efficient XGBoost model, we achieve detection performance (Equal Error Rate (EER) of 17.8\% on 6-channel arrays) comparable to state-of-the-art deep learning baselines, but with a fraction of the computational cost. The proposed system operates with a 1 MB memory footprint and 0.9s latency on a standard ARM Cortex-M7 microcontroller, enabling privacy-preserving, real-time liveness detection on edge devices.
\end{abstract}


\section{Introduction}

Voice authentication is increasingly ubiquitous in smart home and IoT applications, yet it remains vulnerable to replay attacks where an attacker plays back a recording of a legitimate user. To counter this, the research community has developed audio anti-spoofing systems. While recent advancements in Deep Neural Networks (DNNs) have achieved high detection accuracy, they suffer from two critical limitations in edge deployment scenarios.

First, end-to-end deep models operate as ``black boxes'', learning abstract spectro-temporal patterns~\cite{tom2018end,lavrentyeva2017audio} that offer little insight into the decision-making process. The ASVspoof challenge~\cite{liu2023asvspoof} highlighted the dominance of deep models such as RawNet2~\cite{tak2021end} and AASIST~\cite{jung2022aasist}. This opacity creates security blind spots and often leads to overfitting, where models learn to detect specific environmental noise rather than the attack itself. Second, the high parameter count and matrix-multiplication density of these models necessitate powerful hardware (GPUs) or cloud offloading. Cloud processing introduces unacceptable latency for real-time authentication and raises significant privacy concerns regarding the transmission of raw biometric audio. In contrast to existing ``lightweight'' deep learning approaches~\cite{li2025lightweight,ahmed2020void,ma2023lightweight}, our work targets an ultra-low resource tier (microcontrollers), prioritizing interpretability and extreme efficiency without sacrificing robustness in real-world scenarios.

Early audio anti-spoofing systems have explored the use of spatial and physical cues to distinguish live speech from replay attacks. Building on biological principles, Shiota et al. demonstrated the utility of ``pop noise'' as a reliable indicator of liveness, using a tandem detector to isolate these events \cite{shiota2016voice}. This aligns with the concepts discussed by other researchers, who emphasize that human movement and motor control produce unique, dynamic signals that are fundamentally different from the static emission of a loudspeaker~\cite{pezzulo2019body}. Furthermore, another system analyzes sound field dynamics to differentiate the varying mouth aperture of a live speaker from the fixed aperture of a playback device~\cite{yang2023voshield}. These works collectively validate the hypothesis that physical invariants can serve as powerful, interpretable authentication signatures.

In this paper, we build on top of the early methods, and propose to perfect them to lower the existing performance gap with the data-driven DNN-based methods. We also hypothesize that live human speech exhibits unique physical signatures—dynamic source motion, consistent binaural phase cues, and aerodynamic pop noise—that loudspeakers cannot accurately replicate. Unlike a live human, whose natural motor control introduces continuous micro-movements, a loudspeaker is a static point source with rigid phase stability (or electronic jitter). Following first principle research, we formulate novel physical descriptors that quantify \textbf{source kinematics}, \textbf{spatial jitter}, and \textbf{binaural coherence}, and enable building a small and fast lightweight system. Our contributions are:
\begin{itemize}
    \item We introduce a set of manually derived features to differentiate between biological human speech and replayed audio. To the best of our knowledge, this marks the first application of trajectory analysis, such as velocity and acceleration of Time Difference of Arrival (TDoA) estimates for audio anti-spoofing applications.
    \item Recognizing the limitations of deep learning systems, we prioritize a lightweight approach designed for the IoT edge devices. We provide detailed resource estimations for implementation on an ARM Cortex-M7 microcontroller, demonstrating that our method operates within a strict 1 MB memory footprint with an inference latency of approximately 0.9 seconds.
    \item We rigorously evaluate our method against state-of-the-art (SOTA) models and legacy spectral baselines. Our results indicate that while we significantly reduce computational overhead, the proposed method maintains competitive performance and guarantees transparent, physics-based explainability.
\end{itemize}

\section{Methods}
Our system classifies 1-second multi-channel segments $X \in \mathbb{R}^{C \times N}$ as \textit{bona fide} or \textit{spoof}. $C$ is the number of microphone channels and $N$ is the number of samples. From $X$, we extract a physical descriptor vector across three groups capturing acoustic liveness: (i) source kinematics and static spatial descriptors via TDoA trajectories, (ii) binaural coherence and phase dynamics via inter-channel level and phase differences, and (iii) aerodynamic and articulatory descriptors via low-frequency pneumatic energy and spectral-centroid statistics. An XGBoost model classifies the final vector.


\subsection{Source Kinematics \& Static Spatial Descriptors}
\label{first-descriptor}
We model the speaker not as a static coordinate, but as a dynamic system subject to biological motor control. For a selected pair of channels $(i, j)$, we estimate the instantaneous TDoA $\tau_{i,j}(t)$ for frames $t \in \{1, \dots, T\}$ using the Generalized Cross-Correlation with Phase Transform (GCC-PHAT). PHAT whitens the signal and keeps only phase information, reducing reverberation effects, so delay estimation focuses on the direct path. The whitened cross-correlation function is obtained via the Inverse Fast Fourier Transform (IFFT):

\begin{equation}
R_{i,j}^{(phat)}(\Delta \tau, t) = \int_{-\infty}^{\infty} \frac{P_{i,j}(f,t)}{|P_{i,j}(f,t)|} e^{j2\pi f \Delta \tau} df
\end{equation}
where $P_{i,j}(f,t)$ is the cross-power spectrum. The time delay is identified by the peak lag $\tau_{i,j}(t) = \arg \max_{\Delta \tau} R_{i,j}^{(phat)}(\Delta \tau, t)$, confined to a physical aperture $\tau_{max}=1$ms.

We treat the sequence $\tau_{i,j}(t)$ as a trajectory of angular motion. To capture liveness, we compute the motion dynamics, specifically the source velocity $v(t)$ and acceleration $a(t)$:
\begin{align}
    v_{i,j}(t) &= \tau_{i,j}(t) - \tau_{i,j}(t-1) \\
    a_{i,j}(t) &= v_{i,j}(t) - v_{i,j}(t-1)
\end{align}

From these trajectories, we extract \textbf{RMS Velocity} ($V_{rms} = \sqrt{\frac{1}{T}\sum v(t)^2}$) to quantify the overall movement energy, \textbf{Peak} Velocity to identify the largest abrupt shift, and \textbf{Mean Absolute Acceleration} to quantify movement roughness. These metrics distinguish the natural postural sway of humans from the mechanical immobility of loudspeakers.

Complementing these dynamics, we compute \textbf{Static Spatial Descriptors} consisting of the the global mean $\mu_{\tau, i, j}$ and variance $\sigma_{\tau, i, j}^2$ of the raw TDoA values for each pair $(i, j)$. The mean provides a stable estimate of the source's average angular position, while the variance serves as a proxy for spatial focus. Finally, we evaluate \textbf{Spatial Jitter} as a global stability measure by aggregating these per-pair variances across all selected pairs to derive the global mean $\bar{\sigma}_{\tau}^2$ and standard deviation $s_{\sigma_{\tau}^2}$. These global descriptors summarize the overall spatial consistency of the source across the microphone array.

\subsection{Binaural Coherence \& Phase Dynamics}
We analyze the stability of the sound field using Inter-channel Level (ILD) and Phase Differences (IPD) within the speech-dominant band (300–8000 Hz). The audio signals are transformed into the time--frequency domain using STFT, yielding spectra $X_i(f,t)$ for channel~$i$. Then we have:
\begin{align}
\text{ILD}_{i,j}(f, t) &= 20 \log_{10} \left( \frac{|X_i(f, t)| + \epsilon}{|X_j(f, t)| + \epsilon} \right) \\ 
\text{IPD}_{i,j}(f, t) &= \sin(\angle X_i(f, t) - \angle X_j(f, t)),
\label{eq:IPD}
\end{align}
where $\epsilon$ is a small constant to prevent division by zero. We then extract statistical features (mean and variance) across time and frequency.
From each pair's $\text{ILD}_{i,j}(f, t)$ and $\text{IPD}_{i,j}(f, t)$, we compute six descriptors: the \textbf{global Mean ILD/IPD}, the \textbf{ILD/IPD Temporal Variance Mean}, and the \textbf{ILD/IPD Temporal Variance Spread} (Standard Deviation). A coherent point source (human) typically exhibits a focused variance profile, whereas replays often show erratic spectral spread.


To detect the unnatural precision of digital clocks, we compute the \textbf{Phase Velocity Zero-Crossing Rate} (ZCR). Unlike a live human, whose movements and vocal vibrations introduce organic fluctuations, digital hardware plays audio based on a high-precision internal clock. To compute the ZCR, we first average the IPD across frequency bins $\mathcal{F}$ to get a global phase time-series $\overline{\text{IPD}}_{i,j}(t)$. Its temporal derivative, $\phi'_{i,j}(t) = \overline{\text{IPD}}_{i,j}(t) - \overline{\text{IPD}}_{i,j}(t-1)$, represents phase velocity. The ZCR of this velocity is calculated as:
\begin{equation}
\text{ZCR}_{\phi} = \frac{1}{2(T-1)} \sum_{t=2}^{T} |\text{sgn}(\phi'_{i,j}(t)) - \text{sgn}(\phi'_{i,j}(t-1))|.
\end{equation}
High ZCR indicates rapid, electronic phase flipping (jitter), distinct from the smooth phase drifts of biological movement.

\subsection{Aerodynamic \& Articulatory Descriptors}

To quantify articulatory aerodynamics and detect genuine airflow, we define the \textbf{Pop Ratio} ($P_R$), isolating the pneumatic energy of plosives using a 4th-order Butterworth low-pass filter $H_{lp}$ (cutoff 150 Hz):

\begin{equation}
e_c(t) = \frac{\sum |H_{lp} \ast x_c(t)|^2}{\sum |x_c(t)|^2 + \eta}, \quad \overline{e}_c = \frac{1}{T}\sum_{t=1}^{T}e_c(t)
\label{eq:pop_ratio}
\end{equation}
where $\eta$ is a small stabilization constant, and $\overline{e}_c$ represents the time-averaged energy ratio for each audio channel $x_c(t)$. To obtain the final feature representation, we compute the mean and standard deviation of $\overline{e}_c$ across all channels in the array. These act as a ``breath detector''; loudspeakers often fail to reproduce these low-frequency bursts with correct intensity.

Simultaneously, we analyze the Articulatory Dynamics of the source using the \textbf{Spectral Centroid} ($SC$) as a proxy for vocal tract movement. For every frame $t$ and channel $c$, we compute the center of mass of the power spectrum $S_c(f, t)$, 
\begin{equation} 
SC_c(t) = \frac{\sum_{f} f \cdot |S_c(f, t)|}{\sum_{f} |S_c(f, t)| + \eta}.
\label{eq:articulation}
\end{equation}

Complementing this, we analyze the dynamics of the spectral centroid using a hierarchical aggregation process. First, we characterize the temporal evolution for each channel by computing the mean and standard deviation. Second, we perform spatial aggregation by calculating the mean and standard deviation of these temporal metrics across channels. This dual-layer approach captures two physical signatures: Articulatory Liveness, where humans exhibit high temporal variability due to natural phoneme shifts, and Spatial Uniformity, where genuine acoustic propagation maintains consistent spectral properties across all channels. Replayed audio often exhibits spectral smoothing or unnatural consistency that contrasts with the articulatory dynamics of a live speaker.

\subsection{Classifier}

The extracted feature vector is classified using an Extreme Gradient Boosting (XGBoost) model. This gradient boosting framework was selected for its superior performance on tabular physical data and extreme computational efficiency compared to neural networks. 
We re-weighted the binary logistic loss to amplify gradient contributions from the minority class during training.
To improve generalization and prevent overfitting, we incorporated $L_1$ and $L_2$ regularization.


\section{Experiments}

\subsection{Dataset and Evaluation Protocol}
We evaluate our proposed approach on the open-source ReMASC dataset \cite{DBLP:journals/corr/abs-1904-03365}, a diverse multi-channel audio corpus tailored for replay attack detection. The dataset captures recordings across four distinct environments: outdoor (Env-A), indoor (Env-B and C), and a vehicle (Env-D). These acoustic scenes were captured using four microphone arrays (D1–D4) with unique channel topologies. Specifically, D1, D2, and D3 feature 2, 4, and 6 channels, respectively, sampled at 44.1 kHz, whereas D4 integrates 7 channels sampled at 16 kHz. Because our framework relies on physical signatures extracted from microphone pairs, we focus our analysis on the higher-density setups: D3 (a circular array) and D4 (a circular-plus-center array). To ensure a fair consistency with baseline models, we restrict our evaluation to the first second of each recording.


For evaluation, we report Equal Error Rate (EER) under two protocols: i) Standard: Official train/test split in ReMASC, and ii) Environment-Independent (Unseen): A leave-one-environment-out strategy trains on all but one environment and tests on the excluded one to evaluate robustness to unseen acoustics.
%
Finally, the hyperparameters of the XGBoost classifier are optimized using 5-fold stratified cross-validation on the training set.

\subsection{Implementation Details}
\label{feature_details}

Temporal features use a 32 ms frame with a 16 ms hop size. TDoA estimation is strictly constrained to a physical aperture of $\tau_{max} = 1$ ms to reject late wall reflections. To ensure reproducibility, we detail the composition of the concatenated physical feature vectors. The extraction yields a lightweight, highly interpretable footprint of 158 total features for the 6-channel array (D3) and 215 total features for the 7-channel array (D4), broken down as follows:\\
\textbf{Source Kinematics (27 features for D3; 36 for D4)}: Derived from TDoA trajectories. Computing TDoA over all microphone pairs introduces redundant delay information from closely-spaced pairs, inflating the feature dimensionality without adding discriminative power. We therefore restrict the analysis to a uniformly distributed subset (listed in Table~\ref{tab1}), from which we extract RMS velocity, peak velocity, and mean absolute acceleration. This subsampling also reduces computation in line with our lightweight target. The same subset is reused in the Static Spatial Descriptors block.

\textbf{Static Spatial Descriptors (20 features for D3; 26 for D4)}: Includes the local mean and variance of TDoA values for each microphone pair. This category also incorporates global spatial jitter to summarize overall spatial stability.

\textbf{Binaural \& Phase Dynamics (105 features for D3; 147 for D4)}: Computed across all microphone pairs ($C(6,2)=15$ pairs for D3; $C(7,2)=21$ pairs for D4). Each pair contributes exactly 7 statistical metrics: Mean IPD, IPD Temporal Variance Mean, IPD Temporal Variance Spread, Mean ILD, ILD Temporal Variance Mean, ILD Temporal Variance Spread, and Phase Velocity ZCR.\\
\textbf{Aerodynamic \& Articulatory (6 features for both)}: Extracted from all channels and statistically aggregated across frames and channels, encompassing the Pop Ratio ($P_R$) and Spectral Centroid dynamics, which remain independent of the overall array dimension.

\begin{table}[ht]
\centering
\caption{
Configuration of input channels and microphone pairs used for extracting Source Kinematics \& Static Spatial Descriptors (Section \ref{first-descriptor}) across devices.}

\label{tab1}
\begin{tabular}{cc}
\toprule
\textbf{Device} & \textbf{Microphone pairs used in TDoA} \\
\midrule
D3 & (1,3), (1,5), (2,4), (2,6), (3,5), (4,6), (1,4), (2,5), (3,6) \\ 
\multirow{2}{*}{D4} & (1,3), (1,5), (2,4), (2,6), (3,5), (4,6), (1,4), (2,5), \\ 
   & (3,6), (7,1), (7,3), (7,5) \\
\bottomrule
\end{tabular}
\end{table}


Binaural cues (ILD, IPD) are derived using a dedicated STFT with a 75\% overlap Hanning window: 1024 samples (256-sample hop) with a 300--8000~Hz mask for D3, and 512 samples (128-sample hop) with a 300--7500~Hz mask for D4.

\section{Results}

\subsection{Comparison with State-of-the-Art and Edge Feasibility}

Table~\ref{tab:combined_results} compares our method against Neural Network baselines (\textit{NN-*})~\cite{9099075} and the recent attention-based ALRAD model~\cite{neri2025multi}. On the D3 device, our method achieves an EER of 17.8\%, outperforming, on average, the NN-based baselines and performing comparably to the heavier ALRAD architecture. While deep models achieve slightly lower error rates in matched conditions, they do so at a massive computational cost. Our method's performance validates that explicit interpretable physical invariants are discriminative enough to serve as a standalone defense or a lightweight pre-filter.

\begin{table}[ht]
\centering
\caption{Comparison of Performance (EER), Memory usage, and Latency.}
\label{tab:combined_results}
\renewcommand{\arraystretch}{1}
\resizebox{\columnwidth}{!}{
\begin{tabular}{l c c c c c}
\toprule
\textbf{Method} & \textbf{Config} & \textbf{EER (\%) $\downarrow$} & \shortstack{\textbf{Memory} \\ \textbf{(MB)}} & \shortstack{\textbf{Latency} \\ \textbf{(s)}} & \textbf{Feasible} \\
\midrule
NN-Single       & D3 & 23.7 & 30.5 & 2.96 & $\times$ \\
NN-Dummy MC     & D3 & 24.5 & 30.9 & 8.65 & $\times$ \\
NN-Multichannel & D3 & 16.5 & 30.9 & 8.65 & $\times$ \\
ALRAD           & D3 & 19.5 $\pm$ 2.2 & 1.6 & 5.94 & $\times$ \\
M-ALRAD         & D3 & \textbf{10.4 $\pm$ 2.1} & 1.6 & 5.94 & $\times$ \\
\textbf{Proposed} & D3 & 17.8 $\pm$ 0.3 & \textbf{0.6} & \textbf{0.90} & \checkmark \\
\midrule\midrule
NN-Single       & D4 & 27.5 & 30.5 & 1.97 & $\times$ \\
NN-Dummy MC     & D4 & 25.2 & 30.6 & 2.88 & $\times$ \\
NN-Multichannel & D4 & 19.8 & 30.6 & 2.88 & $\times$ \\
ALRAD           & D4 & 21.7 $\pm$ 2.1 & 1.2 & 2.41 & $\times$ \\
M-ALRAD         & D4 & \textbf{14.2 $\pm$ 0.8} & 1.2 & 2.41 & $\times$ \\
\textbf{Proposed} & D4 & 24.0 $\pm$ 0.1 & \textbf{0.6} & \textbf{0.26} & \checkmark \\
\bottomrule
\end{tabular}
}
\end{table}

We estimated resource usage on a high-performance microcontroller, specifically the ARM Cortex-M7 (e.g., NXP i.MX RT1060 or STM32H7 series), using INT16 quantization and SIMD (Single Instruction, Multiple Data) extensions. The hardware benchmarks, also summarized in Table \ref{tab:combined_results}, reveal a critical divergence in deployment viability:
\begin{itemize}
    \item \textbf{Memory}: Deep baselines require $>1.2$ MB (up to 30 MB for older models), exceeding the on-chip SRAM of standard MCUs (typically 1 MB). External RAM usage introduces significant bus latency. Our model requires only 0.6 MB, fitting entirely in fast internal SRAM.
    \item \textbf{Latency}: The baseline models require $\geq 2$ seconds for inference, rendering them unusable for real-time interaction. Our method executes in $\leq$ 0.9 seconds for 1 second of audio, enabling seamless background authentication.
\end{itemize}

\subsection{Interpretability and Feature Importance}

%
To evaluate system interpretability, we examine feature importance via the Gain metric, mapping XGBoost decisions directly to measurable acoustic and physiological phenomena rather than abstract deep embeddings.
The feature importance is quantified using the \textit{Gain} metric, which represents the average improvement in accuracy brought by a specific feature across all trees in the ensemble. To interpret the important features, we map each feature back to its corresponding physical pillar and microphone pair according to the geometry of the arrays. For Device D3 (6-microphone hexagon array) and Device D4 (7-microphone circular and center array), we reconstruct the labels to reflect specific kinematic, binaural, and pneumatic parameters.

\begin{figure*}[t]
    \centering
    \begin{subfigure}[b]{0.47\textwidth}
        \centering
        \includegraphics[width=\textwidth]{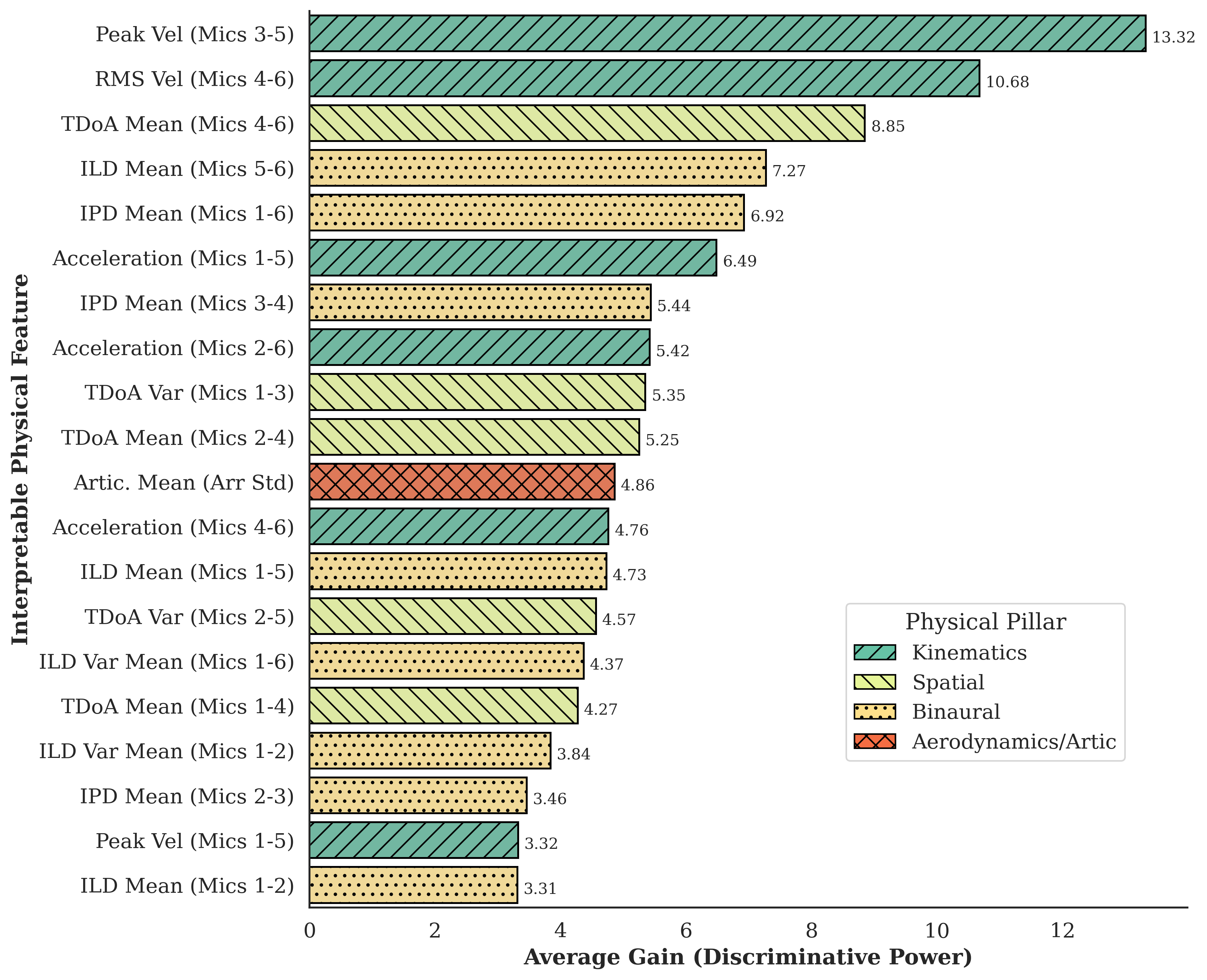}
        \caption{Device D3: 6-microphone hexagonal array.}
        \label{fig:imp_d3}
    \end{subfigure}
    \hfill
    \begin{subfigure}[b]{0.47\textwidth}
        \centering
        \includegraphics[width=\textwidth]{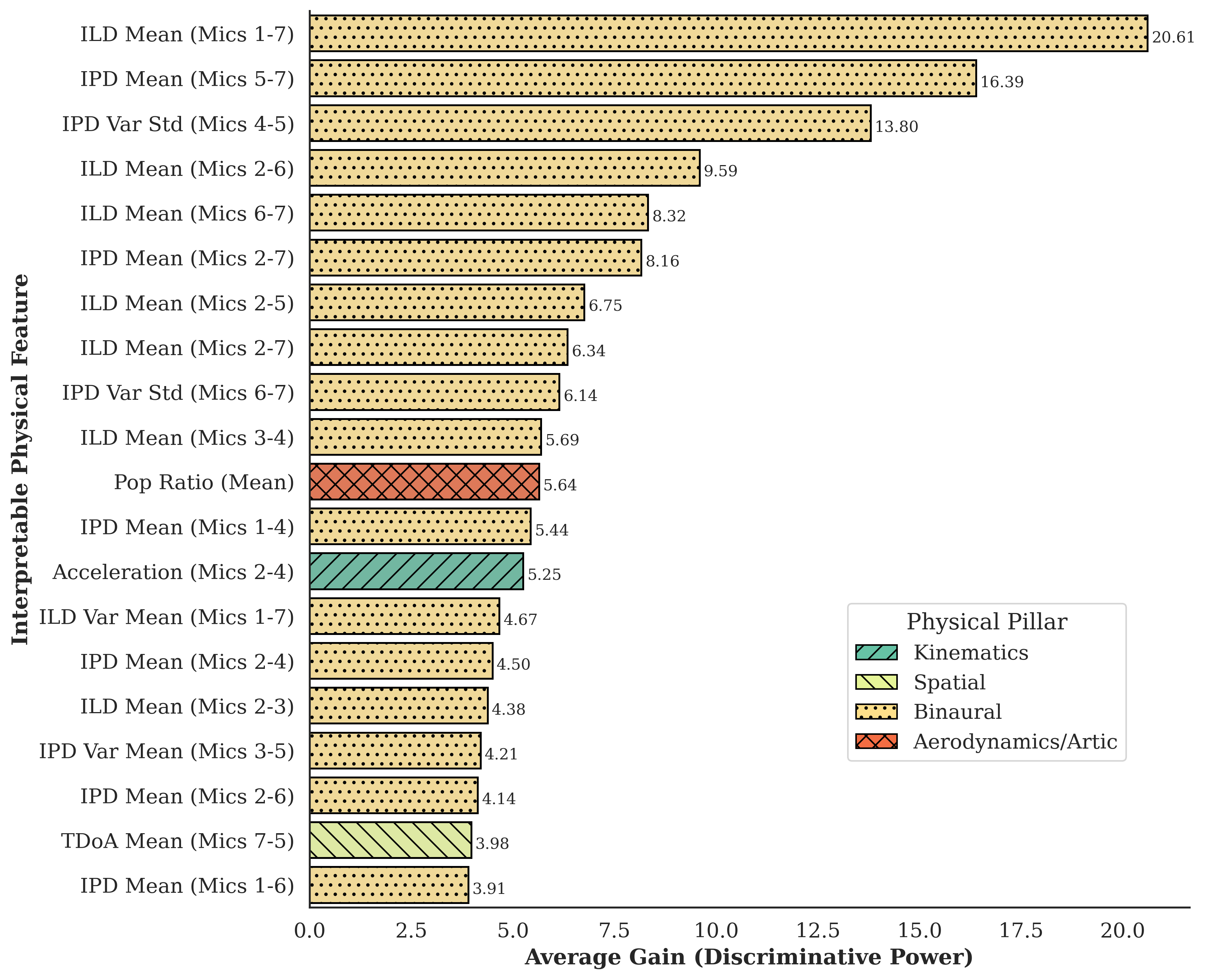}
        \caption{Device D4: 7-microphone circular array with center mic.}
        \label{fig:imp_d4}
    \end{subfigure}
    \caption{Top 20 interpretable physical descriptors ranked by XGBoost Gain. D3 (a) shows reliance on kinematic signatures (biological motion), while D4 (b) leverages the central microphone (Mic 7) to prioritize binaural phase and level dynamics.}
    \label{fig:feature_importance_comparison}
\end{figure*}

The resulting importance rankings, illustrated in Fig.~\ref{fig:imp_d3}  and Fig.~\ref{fig:imp_d4}, confirm that the model's performance is anchored in the physical liveness signatures defined in Section \ref{feature_details}:
\begin{itemize}
    \item \textbf{Device D3 (Hexagonal Mic array)}: The model exhibits a heavy reliance on the Kinematic Signatures of the acoustic source. The top-ranking features, specifically \textit{Peak Velocity} and \textit{RMS Velocity} estimated from microphone pairs with large physical separation, such as Mics 3--5 and 4--6, achieve the highest gain scores.
    This confirms that tracking the ``liveness'' of the source trajectory is the primary discriminator for this topology.
    \item \textbf{Device D4 (Circular + Center Mic array)}: Importance shifts to Binaural Phase features involving the center microphone (Mic 7). The highest importance score is achieved by the \textit{ILD Mean (Mics 1--7)}, which reaches a gain of 20.61. This suggests the model uses the center mic as a reference 
    to capture propagation patterns that differ between live human sources and playback devices.

    \item \textbf{General}: The results across both devices consistently rank features such as the \textit{Pop Ratio} and \textit{Static Spatial Descriptors} among the top physical descriptors. The high ranking of the \textit{Pop Ratio (Mean)} in the D4 configuration validates the effective use of low-frequency pneumatic bursts as a ``breath detector'' for verifying physical liveness. The prominent role of the \textit{Articulatory Mean (Array Std)} confirms that the system successfully utilizes the spatial uniformity and rich articulatory dynamics of genuine acoustic propagation across the microphone array.
\end{itemize}

This alignment between the model's internal logic and the physical hypotheses proves that our system is truly detecting physical presence, providing an inherent safeguard against overfitting to dataset-specific noise. Beyond interpretability, the gain ranking helps optimize resource-constrained deployment. Although the full feature vector satisfies microcontroller memory and latency limits, Fig.~\ref{fig:feature_importance_comparison} indicates a smaller subset suffices due to correlated neighboring microphone pairs. Because XGBoost prioritizes features by gain during training, this ranking enables deployment feature pruning—retaining only the top-$k$ descriptors to reduce memory and inference costs with minimal accuracy degradation.

\subsection{Robustness to Unseen Environments}

To assess the robustness of our approach in realistic, open-set scenarios, we evaluated its performance using an environment-independent protocol, as defined in \cite{DBLP:journals/corr/abs-1904-03365} and \cite{neri2025multi}. In this rigorous testing setup, samples from the target environment are strictly excluded from the training set. For instance, to evaluate performance on the outdoor scenario (Env-A), the model is trained exclusively on Envs B, C, and D. For evaluation, we train a separate model for each device configuration, and the reported results are averaged either across all devices (D1–D4) or restricted to the higher-density arrays (D3 and D4) to analyze the impact of channel count on robustness. 
Table \ref{tab4} compares the performance of our approach against standard baselines and deep learning architectures. The results demonstrate that increasing the microphone array density significantly enhances the model's ability to generalize. While our global average (including all devices) lags behind the state-of-the-art baselines, isolating the performance on D3 and D4 reveals a marked improvement in robustness. Notably, in Env-C, our high-channel configuration (comprising D3 and D4 devices) achieves an EER of 20.5\%, significantly outperforming traditional baselines like CQCC-GMM \cite{DBLP:journals/corr/abs-1904-03365} ($34.6\%$) and deep models like ETECC \cite{PATIL2022101281} ($30.0\%$) and CTECC \cite{9414847} ($32.5\%$). This improvement on D3 and D4 reinforces our hypothesis that physical feature extraction scales positively with channel count, enabling the estimation of increasingly reliable physical parameters.

\begin{table}[ht]
\centering
\caption{Comparison of EER (\%) under the environment-independent (unseen) protocol on the ReMASC dataset.}
\label{tab4}
\renewcommand{\arraystretch}{0.9} 
\setlength{\tabcolsep}{4pt} 
\begin{tabular}{l c c c c}
\toprule
\textbf{Method} & \textbf{Env A} & \textbf{Env B} & \textbf{Env C} & \textbf{Env D} \\
 & (Outdoor) & (Indoor) & (Indoor) & (Vehicle) \\
\midrule
CQCC-GMM \cite{DBLP:journals/corr/abs-1904-03365} & 19.9 & 39.9 & 34.6 & 48.9 \\
ETECC \cite{PATIL2022101281} & 29.0 & 32.5 & 30.0 & 49.9 \\
CTECC \cite{9414847} & 28.5 & 34.7 & 32.5 & 50.0 \\
TECC \cite{9306450} & 26.8 & 35.4 & 31.8 & 50.0 \\
M-ALRAD \cite{neri2025multi} & 13.8 & 20.3 & 15.1 & 24.2 \\
\midrule
\textbf{Proposed Method (D1-4)} & \textbf{37.0} & \textbf{40.7} & \textbf{30.5} & \textbf{48.8} \\
\textbf{Proposed Method (D3-4)} & \textbf{26.0} & \textbf{39.3} & \textbf{20.5} & \textbf{42.2} \\
\bottomrule
\end{tabular}
\end{table}

Our proposed method relies on interpretable physical features that are governed by the laws of physics rather than learned data distributions. This physics-based foundation inherently mitigates the risk of overfitting. While M-ALRAD achieves lower error rates, our method offers a competitive generalization compared to standard baselines (e.g., TECC series) while remaining lightweight and computationally efficient. This balance of speed, interpretability, and generalization makes it uniquely suited for transparent deployment in real-world environments.

\section{Conclusion}

We presented a lightweight audio anti-spoofing system that verifies physical liveness rather than learning black-box patterns. By analyzing source kinematics, binaural coherence, and aerodynamics, we achieve detection performance competitive with deep learning baselines while reducing memory usage by 90\% and latency by nearly 10x. This efficiency unlocks real-time, privacy-preserving replay detection on standard IoT microcontrollers. The system's reliance on multi-channel arrays (ideally 4+) makes it ideal for smart speakers and meeting room hardware. As shown in Table~\ref{tab4}, the proposed method's performance degrades in open spaces (Env. B is an indoor but large room/hall environment). Addressing this limitation, along with extending the framework to moving arrays and highly reverberant open spaces, remains part of future work.


\balance

\bibliographystyle{IEEEtran}
\bibliography{IEEEabrv, references}

\begin{thebibliography}{10}
\providecommand{\url}[1]{#1}
\csname url@samestyle\endcsname
\providecommand{\newblock}{\relax}
\providecommand{\bibinfo}[2]{#2}
\providecommand{\BIBentrySTDinterwordspacing}{\spaceskip=0pt\relax}
\providecommand{\BIBentryALTinterwordstretchfactor}{4}
\providecommand{\BIBentryALTinterwordspacing}{\spaceskip=\fontdimen2\font plus
\BIBentryALTinterwordstretchfactor\fontdimen3\font minus \fontdimen4\font\relax}
\providecommand{\BIBforeignlanguage}[2]{{%
\expandafter\ifx\csname l@#1\endcsname\relax
\typeout{** WARNING: IEEEtran.bst: No hyphenation pattern has been}%
\typeout{** loaded for the language `#1'. Using the pattern for}%
\typeout{** the default language instead.}%
\else
\language=\csname l@#1\endcsname
\fi
#2}}
\providecommand{\BIBdecl}{\relax}
\BIBdecl

\bibitem{tom2018end}
F.~Tom \emph{et~al.}, ``End-to-end audio replay attack detection using deep convolutional networks with attention.'' in \emph{Interspeech}, 2018, pp. 681--685.

\bibitem{lavrentyeva2017audio}
G.~Lavrentyeva \emph{et~al.}, ``Audio replay attack detection with deep learning frameworks.'' in \emph{Interspeech}, 2017, pp. 82--86.

\bibitem{liu2023asvspoof}
X.~Liu \emph{et~al.}, ``Asvspoof 2021: Towards spoofed and deepfake speech detection in the wild,'' \emph{IEEE/ACM Transactions on Audio, Speech, and Language Processing}, vol.~31, pp. 2507--2522, 2023.

\bibitem{tak2021end}
H.~Tak \emph{et~al.}, ``End-to-end anti-spoofing with rawnet2,'' in \emph{IEEE ICASSP}, 2021, pp. 6369--6373.

\bibitem{jung2022aasist}
J.-w. Jung \emph{et~al.}, ``Aasist: Audio anti-spoofing using integrated spectro-temporal graph attention networks,'' in \emph{IEEE ICASSP}, 2022, pp. 6367--6371.

\bibitem{li2025lightweight}
S.~Li \emph{et~al.}, ``A lightweight cross-domain front-end feature extractor for multichannel voice activity and overlapped speech detection,'' in \emph{33rd EUSIPCO}, 2025, pp. 186--190.

\bibitem{ahmed2020void}
M.~E. Ahmed \emph{et~al.}, ``Void: A fast and light voice liveness detection system,'' in \emph{29th USENIX Security Symposium}, 2020, pp. 2685--2702.

\bibitem{ma2023lightweight}
Q.~Ma \emph{et~al.}, ``A lightweight and efficient model for audio anti-spoofing,'' in \emph{ACM MMAsia}, 2023, pp. 1--7.

\bibitem{shiota2016voice}
S.~Shiota \emph{et~al.}, ``Voice liveness detection for speaker verification based on a tandem,'' in \emph{Odyssey: The Speaker and Language Recognition Workshop}.\hskip 1em plus 0.5em minus 0.4em\relax ISCA, 2016, pp. 259--263.

\bibitem{pezzulo2019body}
G.~Pezzulo \emph{et~al.}, ``The body talks: Sensorimotor communication and its brain and kinematic signatures,'' \emph{Physics of life reviews}, vol.~28, pp. 1--21, 2019.

\bibitem{yang2023voshield}
Q.~Yang \emph{et~al.}, ``Voshield: Voice liveness detection with sound field dynamics,'' in \emph{IEEE INFOCOM}, 2023, pp. 1--10.

\bibitem{DBLP:journals/corr/abs-1904-03365}
Y.~Gong \emph{et~al.}, ``Remasc: Realistic replay attack corpus for voice controlled systems,'' \emph{arXiv preprint arXiv:1904.03365}, 2019.

\bibitem{9099075}
Y.~Gong, J.~Yang, and C.~Poellabauer, ``Detecting replay attacks using multi-channel audio: A neural network-based method,'' \emph{IEEE Signal Processing Letters}, vol.~27, pp. 920--924, 2020.

\bibitem{neri2025multi}
M.~Neri and T.~Virtanen, ``Multi-channel replay speech detection using an adaptive learnable beamformer,'' \emph{IEEE Open Journal of Signal Processing}, 2025.

\bibitem{PATIL2022101281}
A.~T. Patil \emph{et~al.}, ``Improving the potential of enhanced teager energy cepstral coefficients (etecc) for replay attack detection,'' \emph{Computer Speech \& Language}, vol.~72, p. 101281, 2022.

\bibitem{9414847}
R.~Acharya \emph{et~al.}, ``Cross-teager energy cepstral coefficients for replay spoof detection on voice assistants,'' in \emph{IEEE ICASSP}, 2021, pp. 6364--6368.

\bibitem{9306450}
H.~Kotta \emph{et~al.}, ``Subband channel selection using teo for replay spoof detection in voice assistants,'' in \emph{APSIPA ASC}, 2020, pp. 538--542.

\end{thebibliography}


\begin{thebibliography}{00}
\bibitem{b1} G. Eason, B. Noble, and I. N. Sneddon, ``On certain integrals of Lipschitz-Hankel type involving products of Bessel functions,'' Phil. Trans. Roy. Soc. London, vol. A247, pp. 529--551, April 1955.
\bibitem{b2} J. Clerk Maxwell, A Treatise on Electricity and Magnetism, 3rd ed., vol. 2. Oxford: Clarendon, 1892, pp.68--73.
\bibitem{b3} I. S. Jacobs and C. P. Bean, ``Fine particles, thin films and exchange anisotropy,'' in Magnetism, vol. III, G. T. Rado and H. Suhl, Eds. New York: Academic, 1963, pp. 271--350.
\bibitem{b4} K. Elissa, ``Title of paper if known,'' unpublished.
\bibitem{b5} R. Nicole, ``Title of paper with only first word capitalized,'' J. Name Stand. Abbrev., in press.
\bibitem{b6} Y. Yorozu, M. Hirano, K. Oka, and Y. Tagawa, ``Electron spectroscopy studies on magneto-optical media and plastic substrate interface,'' IEEE Transl. J. Magn. Japan, vol. 2, pp. 740--741, August 1987 [Digests 9th Annual Conf. Magnetics Japan, p. 301, 1982].
\bibitem{b7} M. Young, The Technical Writer's Handbook. Mill Valley, CA: University Science, 1989.
\end{thebibliography}

\end{document}